# Coherent temporal imaging with analog time-bandwidth compression

Mohammad H. Asghari[1,*] and Bahram Jalali[1,2,3]

[1]*Department of Electrical Engineering, University of California, Los Angeles, CA 90095, USA*
[2]*Department of Bioengineering, University of California, Los Angeles, CA 90095, USA*
[3]*Department of Surgery, David Geffen School of Medicine, University of California, Los Angeles, CA 90095, USA*
*\*Corresponding author: asghari@ucla.edu*

*Abstract—* We introduce the concept of coherent temporal imaging and its combination with the anamorphic stretch transform. The new system can measure both temporal profile of fast waveforms as well as their spectrum in real time and at high-throughput. We show that the combination of coherent detection and warped time-frequency mapping also performs time-bandwidth compression. By reducing the temporal width without sacrificing spectral resolution, it addresses the Big Data problem in real time instruments. The proposed method is the first application of the recently demonstrated Anamorphic Stretch Transform to temporal imaging. Using this method narrow spectral features beyond the spectrometer resolution can be captured. At the same time the output bandwidth and hence the record length is minimized. Coherent detection allows the temporal imaging and dispersive Fourier transform systems to operate in the traditional far field as well as in near field regimes.

## 1. Introduction

Ultrafast non-repetitive phenomena harbor a wealth of fascinating information about a system that is inaccessible to pump-and-probe measurements and to other equivalent-time instruments such as sampling oscilloscopes. Capturing non-repetitive and rare events such as optical rogue waves [1,2] requires real-time instruments. Technical challenges are twofold: (1) digitizing the wideband signal in real time and (2) dealing with the massive volume of data generated in the process.

Coherent Dispersive Fourier Transform (DFT) combines DFT [3,4] and coherent detection. Known as Time Stretch Transform (TST) [5-9] it is used to slow down signals so they can be digitized in real-time. At the same time, coherent detection enables improved sensitivity, digital cancellation of dispersion-induced impairments and optical nonlinearities, and the decoding of phase-modulated optical data formats [7]. By recovering optical amplitude and phase of the time stretched waveform, TST measures both time domain and spectral profile of non-repetitive signals at high-throughput. With complex field detection, DFT can operate in both near-field [5] and far-field [6-9] regimes. However, in the near field there is no Fourier transform, i.e. there is no one-to-one frequency-time mapping. In TST, as well as in temporal imaging, the time–bandwidth product remains constant. For a bandwidth compression of $M$, the record length is increased by $M$ times.

The time stretch technique is inherently an analog optical link but one that uses a broadband laser and large dispersion to slow down the envelope of a fast temporal waveform. Another method to capture ultrafast signals is temporal imaging which duplicates the function of a spatial imaging system in time domain [10-14]. The time lens multiplies (mixes) the signal with a local oscillator that has a linear instantaneous frequency (IF) and this is followed by a Fourier transformation performed by a diffraction grating or temporal dispersion. Conventional temporal imaging can only measure the signal intensity; it does not capture the phase profile.

Recently we introduced a signal transformation that allows capturing fast time waveforms that are beyond the speed of the digitizer and at the same time, minimizing the record length [15-17]. The so-called Anamorphic Stretch Transform (AST) warps the signal with specific phase operator that causes feature-selective time stretch. Upon uniformly sampling the transformed signal, fast features receive a higher sampling density than slow features. The transformation reduces the time bandwidth product and hence the size of digital data produced without losing information. It does so by removing redundancy from the signal in an open loop fashion; i.e. without a-priori knowledge about the signal. AST increases the frame rate and also solves the big data problem that arises during high throughput operation necessary for capturing non-repetitive signals and rare events. To identify the proper AST phase operator that leads to time-bandwidth compression, we introduced a 2D function, Modulation Intensity Distribution (MID) [15-17]. MID can be also used to find the optimum AST phase operator to increase the time bandwidth product in arbitrary waveform generation methods based on frequency to time mapping. AST may also be interpreted in a multitude of ways including warped coherent time-frequency mapping.

In this work, we introduce for the first time the application of Anamorphic Stretch Transform to temporal imaging. The impact is that the spectrum measurement resolution is enhanced while the record length is minimized. In other words, this is a temporal imaging system in which the time-bandwidth product is compressed. Reducing the record length avoids generation of superfluous data and also maximizes the frame rate of spectral measurements. In the proposed method, the signal is warped by mixing it with a local oscillator (LO) that has a nonlinear instantaneous frequency, i.e. an LO with a warped chirp. Coherent detection is then used to measure the complex-field of the output signal. The input signal is reconstructed digitally by back propagation. We use the MID function to design the instantaneous frequency profile that leads to time-bandwidth compression. Compared to conventional temporal imaging, our method results is a shorter record length with the same resolution enhancement factor. Moreover, in contrast to conventional temporal imaging where only the signal time intensity can be captured, the proposed coherent temporal imaging concept can recover complex-field of both time domain and spectrum.

## 2. Principle of operation

Conventional temporal imaging employs a mixer with a linearly chirped local oscillator with temporal profile $h(t)=e^{j \cdot m_2 \cdot t^2/2}$. Here $t$ is the time variable and $m_2$ is the chirp strength factor. We generalize the local oscillator phase profile to an arbitrary warped function of time, $m(t)$, with corresponding instantaneous frequency $IF(t)=\partial[m(t)]/\partial t$ (see Fig. 1(a)). We aim to increase the spectral resolution at the output so it can be captured with a spectrometer with lower resolution. At the same time we minimize the record length to avoid generation of redundant data. Minimizing the record length also maximizes the frame rate of spectral measurement.

To find the optimum instantaneous frequency warp profile, we employ our recently introduced Modulation Intensity distribution (MID). MID is a 2D distribution that describes the output signal intensity as a function of both time and frequency:

$$\text{MID}(\omega_m, t) = \int_{-\infty}^{\infty} E_i(t_1) E_i^*(t_1+t) e^{-j \cdot t \cdot \left[\frac{m(t_1+t)-m(t_1)}{t}\right]} e^{-j \cdot \omega_m \cdot t_1} dt_1 \tag{1}$$

where $\omega_m$ is the envelope frequency. The MID function given by Eq. (1) is the time-domain equivalent of similarly named function introduced in [15-17]. It tells us what chirp profile to use to achieve high spectrum resolution while minimizing the record length. The trajectory at $\omega_m = 0$ in a MID plot (see Fig. 1(b) or Fig. 5) represents the output auto-correlation and the inverse of its width determines the output signal spectral resolution. Also the output signal bandwidth, i.e. the record length, can be measured as the frequency range over which the function has non-zero values. Figure 1(b) is a qualitative MID plot that shows how to engineer the signal time-bandwidth product. The figure compares the MID of the input signal (inset) and the case that it is mixed with a local oscillator with sublinear instantaneous frequency. As seen in Fig. 1(b), the resolution is increased however the bandwidth is not expanded proportionally. In other words, the time-bandwidth product has been compressed. It should be mentioned that the output signal has both amplitude and phase information requiring complex-field detection [18]. The input complex-field spectrum or time domain can then be reconstructed from the measured complex field.

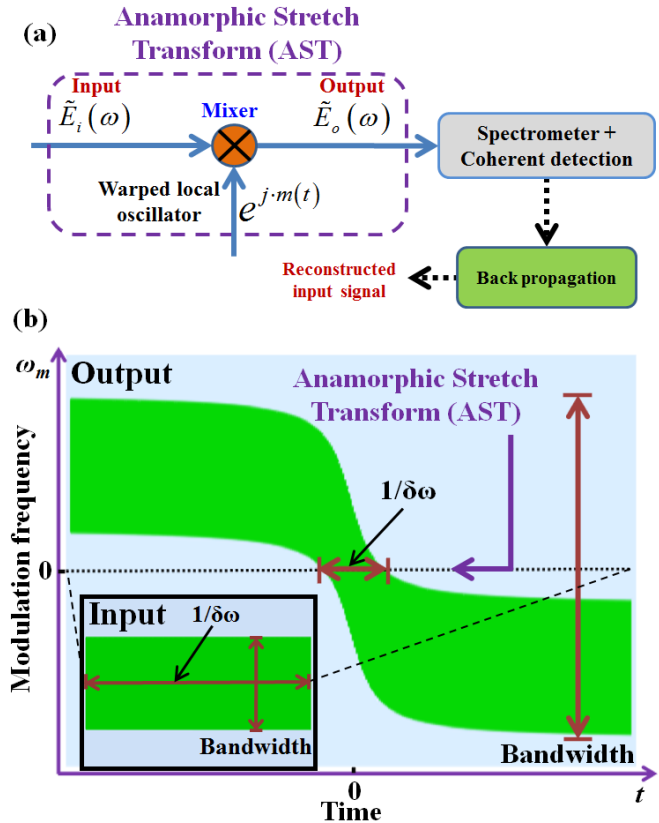

Fig. 1. (a) The proposed warped temporal imaging concept is based on a mathematical transform called Anamorphic Stretch Transform (AST). The transform is implemented using a mixer with a local oscillator with nonlinear instantaneous frequency (IF). The output signal is measured using a spectrometer and coherent detection. The input complex-field waveform is then reconstructed digitally by back propagation. (b) Cartoon showing the application of Modulation Intensity Distribution (MID) to engineer the time-bandwidth product using AST. MID is a 3D plot showing dependence of the envelope intensity (color) on time and envelope frequency. Figure compares the MID of the input signal without any mixer in front and the MID after mixing with a local oscillator with sublinear IF, i.e. AST or warped temporal imaging. Using AST, spectral resolution is increased however the bandwidth is not expanded proportionally, i.e. time-bandwidth compression. MID function is used to design the local oscillator with optimum IF for AST operation.

To compress the TBP, as suggested by MID plots in Fig. 1(b), the system should have a sublinear IF profile. The following function provides a simple mathematical description of such IF function:

$$IF(t) = A \cdot tan^{-1}(B \cdot t), \qquad (2)$$

where $tan^{-1}$ is the inverse tangent function. $A$ and $B$ are arbitrary real numbers. Parameter $A$ determines the amount of output spectral resolution. Parameter $B$ is related to the degree of anamorphism or warping of the instantaneous frequency.

## 3. Numerical Results

As an example on engineering the signal time-bandwidth product using anamorphic temporal imaging, we discuss the optimum IF profile for a local oscillator to enhance the spectral resolution with reduced record length.

To show the utility in single-shot high-throughput spectroscopy, as for the input signal we chose an optical spectrum that resembles spectroscopy traces measured by spectrometers. The resolution of spectral measurement is a key parameter allowing for a precise positioning of the absorption lines. Yet there is a trade-off between the resolution of spectrometer and its update rate. Our method enables fine spectrum features to be captured with a spectrometer that otherwise would not have sufficient resolution. At the same time, our method compresses the time-bandwidth product so the record length is minimized and update rate is maximized.

The optical spectrum under test had ~5.6 THz full bandwidth and 2.8 GHz spectral resolution, i.e. time bandwidth product of 2,000. The input signal spectrum as well as its MID plot are shown in Fig. 2. We intend to reshape the signal so it can be measured using a spectrometer with 50 GHz resolution, i.e. spectral resolution enhancement factor of 17.8.

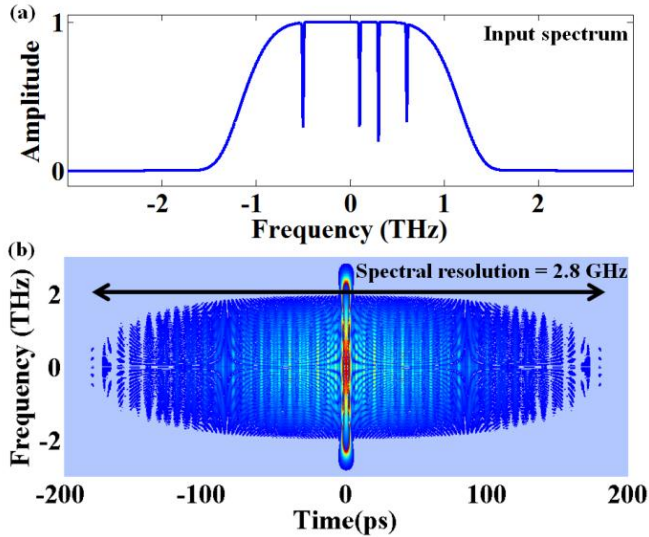

Fig. 2. (a) Input signal spectrum. (b) Modulation Intensity Distribution (MID) of the input signal.

The IF profile for AST operation is chosen such that the output spectral resolution is fixed to 50 GHz, i.e. the target spectral resolution. Parameter $A$ in Eq. (2) was designed to $2.14 \times 10^{13}$ Hz for the given target spectral resolution. To minimize the output bandwidth, i.e. the record length, parameter $B$ in Eq. (2) was designed to $1.05 \times 10^{11}$ Hz. Fig. 3 compares the designed nonlinear IF profile with the linear IF profile with chirp factor of $m_2 = 2.24 \times 10^{24}$ 1/s$^2$ that results in the same spectral resolution.

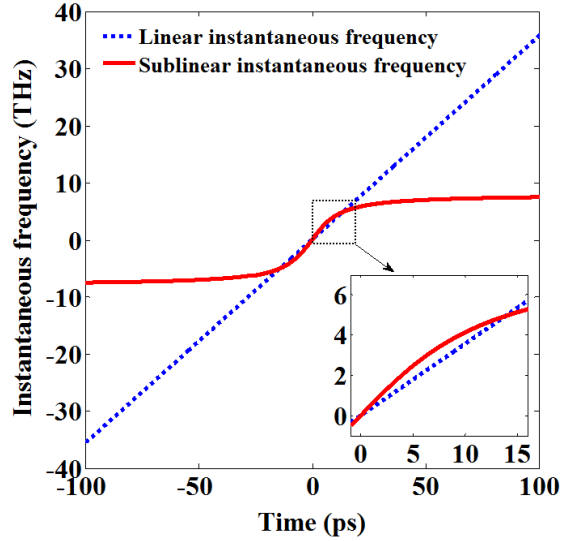

Fig. 3. Comparison of the linear and nonlinear (warped) instantaneous frequency (IF) profiles that result in the same output resolution. As observed in Figs. 4 and 5, while having in the same output spectral resolution the nonlinear IF results in a narrower bandwidth, i.e. a shorter record length.

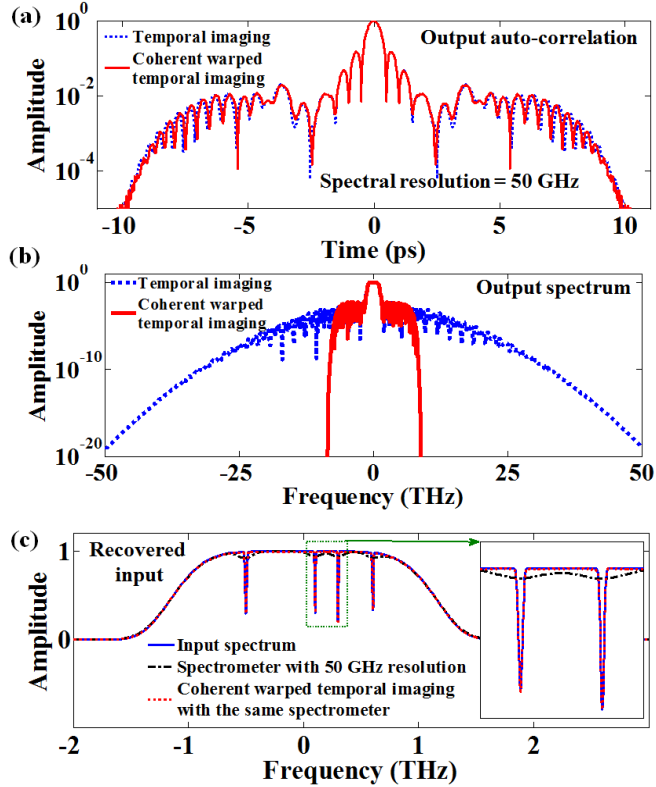

Fig. 4. Time-bandwidth product compression using coherent warped temporal imaging. (a) Comparison of output auto-correlation for the case of linear and nonlinear instantaneous frequency (IF) profiles. Both auto-correlations are reduced to 20 ps, i.e. spectral resolution of 50 GHz. (b) Comparison of output spectrums for linear and nonlinear IF chirp profiles. As seen in (a) and (b) in both cases the spectral resolution is 50 GHz, however the output bandwidth and hence the record length in the case of nonlinear (warped) IF, i.e. AST is 17 THz v.s. 100 THz, i.e. 5.9 times less. (c) Recovered input spectrum using a spectrometer with 50 GHz resolution.

As seen in Fig. 4(a) the output spectral resolution is 50 GHz in both cases. However, the output bandwidth (see Fig. 4(b)) in the case of nonlinear IF is 17 THz vs. 100 THz, i.e. 5.9 times shorter record length using AST. This translates to 5.9 times compression in the time-bandwidth product or 5.9 times higher update rate in spectral measurement. Fig. 4(c) shows the recovered input spectrum using AST method with spectrometer with 50 GHz resolution followed by coherent detection [18]. Original spectrum and the captured spectrum with the same spectrometer but without AST are also shown for comparison. The fine spectrum features are missed using the spectrometer with 50 GHz resolution without AST, however they are fully recovered using AST with the same spectrometer.

Figure 5 compares the MID plots for the cases of linear and warped (nonlinear) IF profiles. These MID plots were used to design and analyze the optimized temporal imaging operation that achieves time-bandwidth compression. The output signal in this case is in the near field regime.

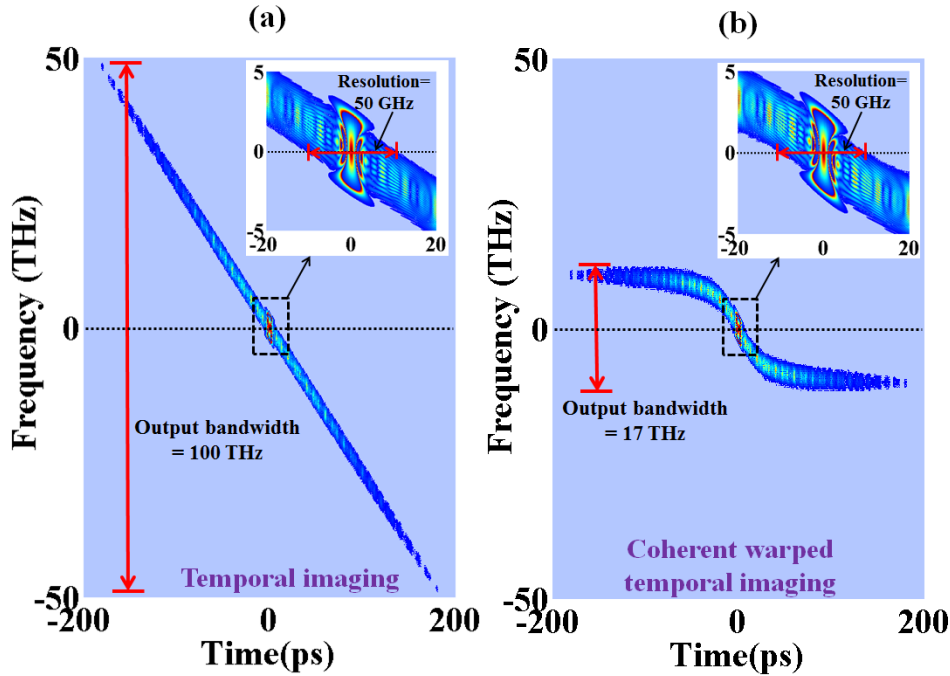

Fig. 5. Left and right figures show the Modulation Intensity Distribution (MID) of the signals in Fig. 4, when the local oscillator has a linear and nonlinear instantaneous frequency (IF) profile, respectively. In both cases the output spectral resolution is 50 GHz, however the output bandwidth in the case of nonlinear IF is 17 THz v.s. 100 THz for linear IF, i.e. 5.9 times decrease in record length. This translates to 5.9 times compression in time-bandwidth product or 5.9 times higher update rate in spectral measurement using Anamorphic Stretch Transform (AST). MID is used to identify the optimum IF profile. The output signal in this case is in near field regime.

## Conclusions

In summary, we introduced for the first time a temporal imaging system that can (i) measure both the temporal profile of fast waveforms as well as their spectrum in real time, and (ii) perform time-bandwidth compression. This coherent warped temporal imaging system is based on mixing the signal with a specific class of warped chirped local oscillators followed by coherent detection. The work represents the application of Anamorphic Stretch Transform to temporal imaging. By removing the redundancy in the input signal, without a-prior knowledge of it, this method increases the frame rate and also solves the big data problem. Big data problem arises in real-time instruments needed for capturing non-repetitive signals and rogue events.